# Stealth Shaper: Reflectivity Optimization as Surface Stylization

KENJI TOJO, The University of Tokyo, Japan
ARIEL SHAMIR, Reichman University, Israel
BERND BICKEL, Institute of Science and Technology Austria, Austria
NOBUYUKI UMETANI, The University of Tokyo, Japan

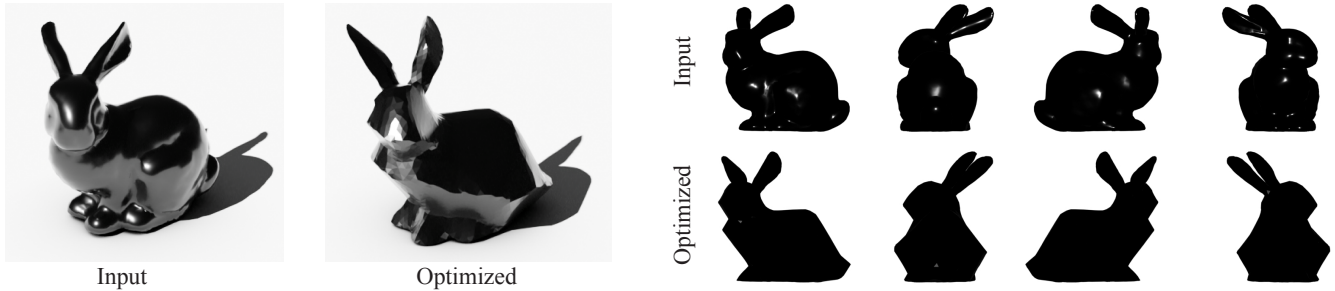

Fig. 1. The input shape (left) is transformed to the stealth one (middle), i.e., the reflection of the shape is minimal in the incoming light direction compared to the original shape (right).

We present a technique to optimize the reflectivity of a surface while preserving its overall shape. The naïve optimization of the mesh vertices using the gradients of reflectivity simulations results in undesirable distortion. In contrast, our robust formulation optimizes the surface normal as an independent variable that bridges the reflectivity term with differential rendering, and the regularization term with as-rigid-as-possible elastic energy. We further adaptively subdivide the input mesh to improve the convergence. Consequently, our method can minimize the retroreflectivity of a wide range of input shapes, resulting in sharply creased shapes ubiquitous among stealth aircraft and Sci-Fi vehicles. Furthermore, by changing the reward for the direction of the outgoing light directions, our method can be applied to other reflectivity design tasks, such as the optimization of architectural walls to concentrate light in a specific region. We have tested the proposed method using light-transport simulations and real-world 3D-printed objects.

## 1 INTRODUCTION

Physical functionalities are often the origin of various features that determine the shapes of manufactured objects. As the architect Luis Sullivan put it, "form follows function." Among such functionalities, the reflectivity of a surface poses interesting design challenges, such as the shapes of reflectors [Patow and Pueyo 2005] and caustic panels [Papas et al. 2011]. Reflectivity is also an important factor in architectural design. For instance, certain building designs may inadvertently cause undesirable reflective glares [Zhu et al. 2019], and reflective walls can be utilized to distribute sunlight [Sakai and Iyota 2017]. Furthermore, the balance between reflectivity and other design objectives can lead to a unique geometric style: the *stealth* design [Knott et al. 2004] minimizes the retroreflectivity of shapes to hide them from radars, resulting in the unique sharply creased shapes commonly observed in fighter aircraft and battleships (Fig. 2). Despite its many applications, however, reflectivity-aware design typically requires specialized knowledge and experimentation, hindering automatic methods and casual 3D modeling.

Physically-based light transport algorithms [Veach 1998] can simulate complex reflections of light on arbitrary 3D models. However, even with a faithful simulator, editing a surface using a 3D tool to improve reflective functionality is a challenging inverse design that typically involves tedious trial-and-error cycles. Recent studies have presented methods for differentiating the light-transport simulations to efficiently solve inverse problems via gradient descent [Li et al. 2018; Nimier-David et al. 2020, 2019]. Although these methods have successfully reconstructed accurate 3D models from images, directly updating shapes using gradients often introduces significant distortion, which is unsuitable for the exploration of functional shapes while maintaining the original geometry.

In this paper, we present a framework for deforming a surface while balancing reflectivity optimization and shape preservation. Inspired by recent geometric stylization approaches [liu and Jacobson 2019; Liu and Jacobson 2021], our method integrates the as-rigid-as-possible (ARAP) shape regularization into the gradient-based reflectivity optimization using a surface-normal-based shape representation. Consequently, our framework allows users to optimize the geometry within a near-isometric deformation of the original shape, thereby preserving the overall geometry, as well as its surface properties, such as mesh quality and texture.

Authors' addresses: Kenji Tojo, The University of Tokyo, Japan, knjtojo@g.ecc.u-tokyo.ac.jp; Ariel Shamir, Reichman University, Israel, arik@runi.ac.il; Bernd Bickel, Institute of Science and Technology Austria, Austria, bernd.bickel@ist.ac.at; Nobuyuki Umetani, The University of Tokyo, Japan, n.umetani@gmail.com.

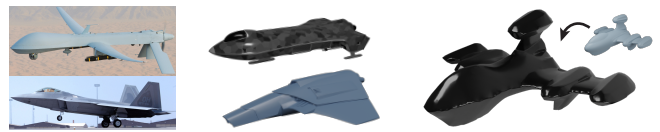

Fig. 2. (Left to right) Real-world stealth aircraft, imaginary 3D spaceships adopting the stealth style, and a *stealth* rocket computed using our reflection optimization. The images are from wikipedia.com in the public domain.

We evaluate our approach by optimizing the stealth properties of 3D models. Specifically, we formulate an energy that evaluates the retroreflectivity of a surface, and minimizes it using the proposed method. The results were fabricated using a 3D printer, and real-world experiments indicate that our method successfully minimizes retroreflection. We also illustrate architectural applications, such as optimizing a building shape to maximize reflection on the street. Lastly, we compare our method with baselines that directly update vertices in gradient descent. We also compare major design choices for adaptive mesh subdivision to overcome the limitations of the ARAP-based deformation and for normal denoising to alleviate stochastic noise arising from the Monte Carlo gradient estimation.

## 2 RELATED WORK

*Appearance-based surface optimization.* A variety of optical effects have been produced via surface optimization, such as bas-reliefs [Alexa and Matusik 2010; Schüller et al. 2014; Weyrich et al. 2007] and meta-materials [Auzinger et al. 2018; Levin et al. 2013; Weyrich et al. 2009]. Existing approaches most closely resembling our own include caustics design [Finckh et al. 2010; Papas et al. 2011; Schwartzburg et al. 2014] and reflector design [Patow and Pueyo 2005], which optimize reflective or refractive light transport on surfaces. Although these methods have successfully produced intricate lighting effects, they typically employ restrictive geometric representations, such as height fields and parametric surfaces [Papas et al. 2011; Patow and Pueyo 2005], to simplify the optimization. In contrast, we employ triangle meshes, which are commonly used as geometric representations in many modeling tools. The prior reflection design approaches are not directly suited for the optimization of polygonal meshes, as the meshes can represent a wider variety of shapes and allow a significantly more flexible deformation controlled by a large number of independent vertices. As demonstrated, our method efficiently optimizes the reflectivity of meshes with around 10K faces, while preserving their original geometry.

Micro-scale surface structures have been explored for the fabrication of various functional materials [Auzinger et al. 2018; Levin et al. 2013; Weyrich et al. 2009]. Similarly, a given reflectivity objective can be achieved using high-frequency surface structures without modifying the shape at the macroscale level. These approaches are orthogonal to ours, as we focus on piecewise smooth surfaces that are easier to fabricate and independent of the mesh resolution.

Although methods for stealth airplane design using gradient-based optimization have been developed in the aerodynamics engineering literature [Li et al. 2019; Zhou et al. 2020], they typically assume a parametric surface representation for specific airplane wing shapes. In addition, they use costly *offline* physical models encompassing wave optics and aerodynamic constraints. Our method relies on a simpler geometric optics model simulated with path tracing, enabling it to support a wider variety of 3D models as inputs, as well as *interactive* modeling scenarios.

*Differentiable rendering.* Recent studies in computer graphics and vision have often used derivatives of rendering algorithms to address inverse problems via gradient descent [Kato et al. 2018; Laine et al. 2020; Liu et al. 2019]. Among them, physically-based methods represent algorithms for evaluating the derivatives of light transport simulations [Gkioulekas et al. 2013; Li et al. 2018; Nimier-David et al. 2020; Zhang et al. 2019] with respect to scene parameters involving geometry, material, and lighting. Our method relies on physically-based differentiable rendering to update the face normals of a mesh, such that they minimize the energy defined in terms of the surface radiance.

Gradient-based inverse rendering of geometry is a challenging task, as the gradients are typically noisy and localized, which may easily collapse the mesh during optimization [Luan et al. 2021; Nicolet et al. 2021]. Nicolet et al. [2021] recently presented a specialized preconditioning scheme based on the mesh Laplacian, drastically improving the robustness of inverse shape reconstruction. However, their method incurs significant destructive deformation in the first several descent steps, which is undesirable when seeking to preserve the original geometry. Unlike these methods, our method is tailored to shape preservation in reflectivity design, and we compare our approach with the Laplacian preconditioning-based method. Note that our ARAP-based regularization can also potentially benefit the shape reconstruction, when, for example, prior knowledge of the target shape is available.

Our work shares a similar motivation to [Liu et al. 2018], wherein differentiable rendering was used to augment geometric modeling. However, they focused on translating the visual effects of image filters into 3D shapes, and did not consider physical light transport. In addition, their method requires frequent remeshing during the optimization, as the mesh vertices are updated directly. In contrast, our method can optimize the real-world reflectivity of a shape, while maintaining the mesh quality using near-isometric deformation.

*Energy-based surface modeling.* Many geometry processing algorithms develop a surface by iteratively minimizing the prescribed surface energy [Brakke 1992]. Such *surface flow* procedures have been applied to a variety of tasks, such as surface fairing [Desbrun et al. 1999; Taubin 1995], conformal deformation [Crane et al. 2013], architectural modeling [Pellis et al. 2019], and morphological operations [Sellán et al. 2020]. Robustness has been a central concern in these methods, and many studies have suggested the use of implicit schemes [Crane et al. 2013; Desbrun et al. 1999; Kazhdan et al. 2012] instead of directly updating the vertices. Although such traditional surface energies are typically derived from local geometries (e.g., curvature), our reflectivity energy relies on light-transport simulations that can handle global effects, such as inter-reflections. As a result, our framework can support various reflectivity objectives, unlike existing shape-from-shading approaches tailored to specific material and lighting conditions [Tosun et al. 2007].

The ARAP energy [Igarashi et al. 2005; Sorkine and Alexa 2007] encourages strong shape preservation and, in conjunction with surface-normal constraints, has been applied to various geometry processing tasks, such as polycube deformation [Zhao et al. 2017], geometric stylization [liu and Jacobson 2019], and developable surface design [Binninger et al. 2021]. Drawing inspiration from these approaches, we formulate an energy function that combines the ARAP-based shape regularization with a reflectivity objective, defined in terms of surface normals. Consequently, our method robustly optimizes the surface reflectivity based on differentiable light-transport simulations with minimal shape distortion.

*Mesh denoising.* Because we estimate the derivative of our surface reflectivity energy using the Monte Carlo method, our approach can benefit from denoising algorithms on 3D surfaces [He and Schaefer 2013; Sun et al. 2007; Zhang et al. 2015b,a]. These methods remove high-frequency noise from properties defined on the surface, such as normals, while preserving the sharp features. We use a method based on total-variation regularization [Zhang et al. 2015b], which yields high-quality results with a low runtime.

## 3 BACKGROUND

This section briefly reviews the existing tools used as a foundation for our energy formulation and shape deformation methods.

### 3.1 Radiometry and Path Tracing

The radiance $L(\mathbf{p}, \boldsymbol{\omega}_o)$ is the light intensity at location $\mathbf{p} \in \mathbb{R}^3$ emanating in the direction $\boldsymbol{\omega}_o \in \mathcal{S}^2$. In physically-based rendering [Pharr et al. 2016], the radiance in a 3D scene is computed using Monte Carlo path tracing, which estimates the recursive integral [Kajiya 1986]

$$L(\mathbf{p}, \boldsymbol{\omega}_o) = \int_{\mathcal{S}^2} f(\boldsymbol{\omega}_o, \boldsymbol{\omega}_i) L(\mathbf{q}, -\boldsymbol{\omega}_i) |\cos \theta_i| \mathrm{d}\boldsymbol{\omega}_i, \quad (1)$$

where $f$ is the BRDF, $\mathbf{q}$ is the intersection of the ray $\mathbf{p} + t\boldsymbol{\omega}_i$ with the surface, and $\theta_i$ is the angle between $\boldsymbol{\omega}_i$ and the normal at $\mathbf{p}$.

*Phong BRDF.* The present study focuses on the Phong reflectance model [Phong 1975], which models a near-specular reflection using a simple analytic formula and offers an efficient importance sampling method [Lafortune and Willems 1994]. Mathematically, the Phong BRDF can be expressed as

$$f(\boldsymbol{\omega}_o, \boldsymbol{\omega}_i) = k_\mathrm{d} \frac{1}{\pi} + k_\mathrm{s} \frac{n+2}{2\pi} \cos^n \alpha, \quad (2)$$

where $\alpha$ is the angle between $\boldsymbol{\omega}_i$ and the perfect mirror reflection direction; and $k_\mathrm{d}$, $k_\mathrm{s}$, and $n$ are parameters that controls the shape of the BRDF [Lafortune and Willems 1994]. We use $n = 30$ throughout our experiments. Although we assume a constant $f$ over the surface, our method could be easily extended to spatially-varying BRDFs.

### 3.2 Derivative of Light Transport

Our method is based on the gradient-based optimization of surface energy defined in terms of radiance. We evaluate the derivative using the reverse-mode differentiation method for light-transport simulations [Nimier-David et al. 2020]. This method, known as radiative backpropagation, allows us to efficiently differentiate a large number of parameters. To obtain the gradient descent algorithm, we analytically derive the derivative of the Phong BRDF (2) and use the unbiased version of the radiative backpropagation.

### 3.3 Normal-Driven Geometric Stylization

Given a 3D mesh with vertex positions $\mathcal{V} \in \mathbb{R}^{N \times 3}$, the *cubic* stylization [liu and Jacobson 2019] determines the vertices $\mathcal{V}'$ of the stylized mesh that preserves the original geometry. Specifically, the stylization minimizes the energy consisting of $L1$ regularization on the surface normals and the ARAP term [Igarashi et al. 2005; Sorkine and Alexa 2007]. This technique has more recently been extended to arbitrary predefined target normals expressed via Gauss maps [Kohlbrenner et al. 2021; Liu and Jacobson 2021]. In essence, this normal-driven stylization minimizes the energy

$$E_\text{style} = \underbrace{\sum_k \sum_{(i,j) \in \mathcal{N}_k} \|R_k \mathbf{e}_{i,j} - \mathbf{e}'_{i,j}\|_2^2}_{E_\text{ARAP}(\mathcal{V}', \mathcal{R})} + \underbrace{\lambda \|R_k \mathbf{n}_k - \mathbf{t}_k\|_2^2}_{E_\text{normal}(\mathcal{R}; \mathcal{T})}, \quad (3)$$

where $\mathcal{N}_k$ is an element defining the vertex neighbourhood, $\mathbf{e}_{i,j}$ and $\mathbf{e}'_{i,j}$ are the edge vectors as $\mathbf{e}_{i,j} = \mathbf{v}_i - \mathbf{v}_j$, $R_k$ is a rotation matrix, $\mathbf{n}_k$ is the initial normal of the $k$-th element, and $\mathbf{t}_k$ are the target directions encoding the geometric style. The target normals $\mathbf{t}_k$ are typically specified as $\mathbf{t}_k = \mathbf{t}(\mathbf{n}'_k)$ using a pre-defined spherical direction map. Although this fixed directional mapping can produce various *geometric* stylization effects, it is unclear how to achieve the stylization as defined by the *physics* of reflection. Our method optimizes the target directions in tandem with the optimization of (3), thereby optimizing the light-transport function while preserving the overall input shape. Further details of the stylization energy are discussed in [liu and Jacobson 2019], including the cotangent weight for the ARAP term.

## 4 METHOD

Our workflow takes as input a surface $\mathcal{M}$ represented as a manifold mesh $(\mathcal{V}, \mathcal{F})$ with the three-dimensional positions $\mathcal{V}$ and triangular faces $\mathcal{F}$. The input surface is then deformed to improve the specified reflective functionality while retaining the original geometry. This process is iterated until the deformation reaches an equilibrium between the reflectivity objective and the shape preservation. We write $\mathcal{V}$ as the original reference vertex positions and $\mathcal{V}'$ as the deformed vertex positions.

Deformation is computed as the minimizer of the energy consisting of reflectivity and geometric regularization terms, and robustly optimized through a tailored gradient-based minimization scheme. For the regularization, we use the ARAP energy owing to its effectiveness in shape preservation, as demonstrated in many prior studies [Igarashi et al. 2005; Sorkine and Alexa 2007]. Consequently, our energy can be expressed as

$$E_\text{total} = E_\text{refl}(\mathcal{V}') + E_\text{ARAP}(\mathcal{V}', \mathcal{R})$$

where $E_\text{refl}$ is the reflectivity energy based on the surface radiance, for which only the first-order gradient is available for optimization. The energy could be minimized using alternating optimization [Li and Lin 2015; Parikh and Boyd 2014] by minimizing the ARAP term based on the standard local-global approach [Sorkine and Alexa 2007], and the reflectivity term using gradient descent, as

$$\mathcal{V}' \leftarrow \mathcal{V}' - \eta \frac{\partial E_\text{refl}}{\partial \mathcal{V}'}.$$

However, updating a mesh with vertex displacements is a delicate task that can easily introduce irreparable artifacts, particularly when stepping $\eta$ is large [Desbrun et al. 1999]. This problem is exacerbated in our setting by the stochastic noise arising from the Monte Carlo integration in the radiance derivative estimation. Although the bi-Laplacian preconditioning [Nicolet et al. 2021] can protect the mesh from artifacts, it causes a large irrelevant deformation in the first several descent steps. As a result, the optimization often results in

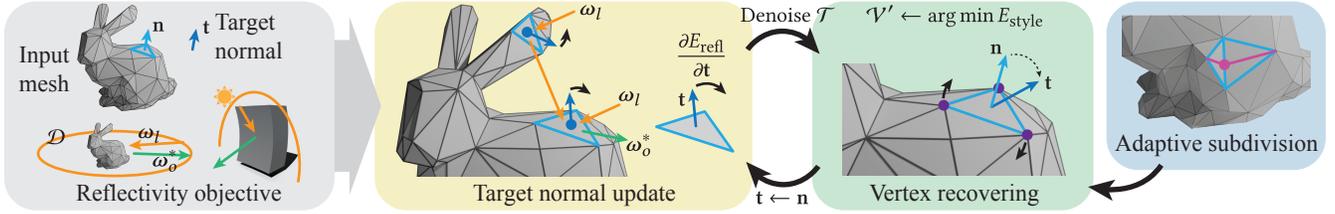

Fig. 3. Workflow of our method. We optimize the reflectivity of the input shape by iterating the gradient-based normal optimization and ARAP-based vertex recovering. We also perform an adaptive edge subdivision to introduce sharp creases, crucial for certain reflectivity properties, such as the stealth property.

**ALGORITHM 1:** Reflective surface optimization

   **input** $(\mathcal{V}, \mathcal{F})$
   $\mathcal{V}' \leftarrow \mathcal{V}$
   **for** $1, 2, \ldots, n_{\text{vertex}}$ **do**
      $\mathcal{T} \leftarrow \text{FaceNormal}(\mathcal{V}', \mathcal{F})$
      **for** $1, 2, \ldots, n_{\text{gradient}}$ **do**
         $\mathcal{G} \leftarrow \partial E_{\text{refl}}/\partial \mathcal{T}$
         $\mathcal{T} \leftarrow \mathcal{T} - \eta \mathcal{G}$
      $\mathcal{T} \leftarrow \text{Denoise}(\mathcal{T})$
      $\mathcal{V}' \leftarrow \arg\min_{\mathcal{V}', \mathcal{R}} E_{\text{style}}(\mathcal{V}', \mathcal{R}; \mathcal{T})$
   **return** $(\mathcal{V}', \mathcal{F})$

a shape that significantly deviates from the original or in a suboptimal shape.

Inspired by the recent normal-driven geometric stylization [liu and Jacobson 2019], we overcome this challenge by introducing auxiliary variables $\mathcal{T} \in \mathbb{R}^{M \times 3}$ for the target face normals. Thus, we rewrite the energy as

$$E_{\text{total}} = E_{\text{refl}}(\mathcal{T}) + E_{\text{style}}(\mathcal{V}', \mathcal{R}; \mathcal{T}). \quad (4)$$

As shown in Algorithm 1, we minimize (4) by alternating the minimization of the two terms. This allows us to optimize $E_{\text{refl}}$ through gradient descent, while updating the vertices $\mathcal{V}'$ based only on the quadratic energy $E_{\text{style}}$, leading to a significantly more robust deformation. The remainder of this section describes the reflectivity energy and its derivative estimation. We then introduce methods to further improve the results, such as normal denoising and adaptive edge splitting, and briefly discuss implementation details. See Fig. 3 for a workflow diagram.

### 4.1 Surface Reflectivity Energy

In this study, We focus on a directional emitter (i.e., parallel rays of light) with incident light in the direction $\omega_l$. Such directional emitters approximate sunlight, and are also observed under various artificial lighting conditions. Given a lighting direction $\omega_l$, we consider the problem of reflecting or deflecting the light in the target direction $\omega_o^*(\mathbf{p})$ defined at each point $\mathbf{p}$ on the surface. Assuming the single target direction makes the problem feasible to handle, but still accommodates many applications, as we demonstrate later. Thus, our reflectivity energy is expressed via the L2 norm between the target $L^*$ and the current radiance values as

$$E^{\omega_l}(\mathcal{M}) = \frac{1}{2} \int_{\mathcal{M}} \left\| L^* - L(\mathbf{p}, \omega_o^*; \omega_l) \right\|^2 V(\mathbf{p}, \omega_o^*) \mathrm{d}A, \quad (5)$$

where $V(\mathbf{p}, \omega)$ is the binary visibility with $V(\mathbf{p}, \omega) = 0$ for the direction $\omega$ blocked by $\mathcal{M}$, and 1 otherwise.

Because practical problems often encompass multiple lighting directions simultaneously (e.g., the sun moving in the sky or the range of radar directions in stealth design), we define our final radiance-based energy by averaging (5) over the set $\mathcal{D} \subseteq \mathcal{S}^2$ of the possible lighting directions:

$$E(\mathcal{M}) = \frac{1}{|\mathcal{D}|} \int_{\mathcal{D}} E^{\omega_l}(\mathcal{M}) \mathrm{d}\omega_l, \quad (6)$$

where $|\mathcal{D}|$ denotes the area of $\mathcal{D}$.

*The stealth energy.* We demonstrate our framework using several energy functions derived from our formulation. The stealth design problem involves reducing the object's retro-reflectivity when observed from a circular band of possible radar directions. This can be framed using the reflectivity energy by setting $L^* \equiv 0$ and $\omega_o^* \equiv \omega_l$. Thus, the stealth objective function for a single lighting direction can be expressed as

$$E^{\omega_l}_{\text{stealth}}(\mathcal{M}) = \frac{1}{2} \int_{\mathcal{M}} L(\mathbf{p}, \omega_l; \omega_l)^2 V(\mathbf{p}, \omega_l) \mathrm{d}A, \quad (7)$$

and the set of directions is defined as $\mathcal{D}_{\theta_0} := \{(\theta, \phi) \in [-\theta_0, \theta_0] \times [0, 2\pi]\}$ for $\theta_0 > 0$ (we set $\theta_0 = 20°$ in our experiments).

*Other reflectivity energies.* Although our framework is primarily evaluated using the stealth energy, our reflectivity energy can express other interesting functions. For example, we can maximize the reflection by alternatively setting $L^*$ to a large constant, or using a loss function $1/(L + \epsilon)$ in (5), rather than the L2 norm (we use the latter approach in our experiment). We can also direct reflection toward (or away from) the subset of three-dimensional space (e.g., points, lines, and planes). To this end, the target direction $\omega_o^*(\mathbf{p})$ is set to the direction to the closest point in the subset. For example, we concentrate the reflection to a line connecting points $\mathbf{q}_1$ and $\mathbf{q}_2$ by setting $\omega_o^*(\mathbf{p}) = (\mathbf{q}_1 - \mathbf{p}) \cdot (\mathbf{q}_2 - \mathbf{q}_1)/(\|\mathbf{q}_1 - \mathbf{p}\|\|\mathbf{q}_2 - \mathbf{q}_1\|)$. We demonstrate several such energies in our results. Although the remainder of this section discusses the stealth energy, the derivation applies to such other energies with minimal adaptation.

### 4.2 Reflectivity Derivative Estimation

We use the gradient of the stealth energy to update the target normals $\mathcal{T}$ such that they reduce the energy value. Therefore, it is convenient to discretize the energy (7) on the faces as

$$E^{\omega_l} \approx \frac{1}{2} \sum_{k \in \mathcal{F}} A_k L(\mathbf{c}_k, \omega_l; \omega_l)^2 V(\mathbf{c}_k, \omega_l). \quad (8)$$

Here, $A_k$ is the area of the $k$-th triangular face, and $\mathbf{c}_k$ is a random point on the face. Thus, the final energy $E$ can be computed as

$$E \approx \frac{1}{|\mathcal{D}|} \int_{\mathcal{D}} E^{\omega_l} d\omega_l = \frac{1}{2} \sum_{k \in \mathcal{F}} A_k \frac{1}{|\mathcal{D}|} \int_{\mathcal{D}} Q_k(\omega_l) d\omega_l,$$

where $Q_k(\omega_l) := L(\mathbf{c}_k, \omega_l; \omega_l)^2 V(\mathbf{c}_k, \omega_l)$.

Our method relies on the derivative of the energy $\partial E / \partial \mathcal{T}$. As light transport involves discontinuous visibility, the differential and the integral operations do not exchange, which produces an additional boundary gradient term that requires dedicated algorithms for accurate integration [Bangaru et al. 2020; Li et al. 2018; Loubet et al. 2019]. Because our method recovers vertex positions by minimizing the energy $E_{\text{style}}$, vertex displacements for a given normal perturbation, which is typically required to evaluate the boundary term, are not straightforward to predict. Although the boundary term is crucial in geometric reconstruction using inverse rendering to *follow* the object's contours, our objective is to *maintain* the object's contours to preserve the shape. Therefore, we simply ignore the boundary term, focusing on the continuous component of light transport. We leave the discontinuous component for future work.

In summary, we compute the energy gradient as

$$\frac{\partial E}{\partial \mathcal{T}} \approx \frac{1}{2} \sum_{k \in \mathcal{F}} A_k \frac{1}{|\mathcal{D}|} \int_{\mathcal{D}} \frac{\partial Q_k}{\partial \mathcal{T}} d\omega_l, \quad (9)$$

where

$$\frac{\partial Q_k}{\partial \mathcal{T}} = L(\mathbf{c}_k, \omega_l; \omega_l) \frac{\partial}{\partial \mathcal{T}} L(\mathbf{c}_k, \omega_l; \omega_l) V(\mathbf{c}_k, \omega_l).$$

To evaluate the integral $1/|\mathcal{D}| \int_{\mathcal{D}} \partial Q_k(\omega_l)/\partial \mathcal{T} d\omega_l$, we sample $N_{\text{dir}}$ directions $\omega_l^i$ and calculate the finite sum $1/N_{\text{dir}} \sum_i \partial Q_k(\omega_l^i)/\partial \mathcal{T}$. We then evaluate the radiance $L$ and the derivative $\partial L/\partial \mathcal{T}$ in each $\partial Q_k/\partial \mathcal{T}$ via ordinary path tracing and the radiative backpropagation [Nimier-David et al. 2020], respectively.

### 4.3 Denoising Normals

As our gradient evaluation relies on the Monte Carlo method, the normal directions $\mathcal{T}$ following the descent steps may contain noise, naturally introducing undesirable high-frequency bumps on the surface through vertex updates. Although such noise can be eliminated by using a large number of samples, this approach is often too costly in interactive applications. Instead, we independently take a relatively small number of samples at each face and locally average the gradient effect by smoothing the normals $\mathcal{T}$ once before updating the vertices.

As suggested by real-world stealth designs, we found that sharp geometric features are sometimes crucial for reducing the reflectivity energy. Therefore, the denoising step of our method shares a similar motivation with classical mesh denoising algorithms to filter high-frequency bumps while preserving the major geometric features. Accordingly, we apply the normal filtering algorithm based on total variation regularization [Zhang et al. 2015b] to efficiently produce feature-preserving denoising results by iterating a sparse linear solve and a closed-form update.

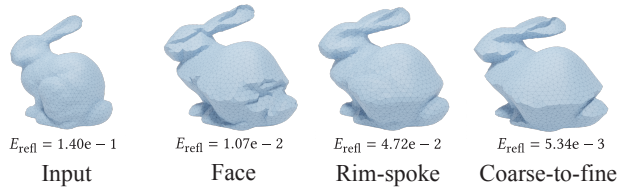

$E_{\text{refl}} = 1.40\text{e}-1$  $E_{\text{refl}} = 1.07\text{e}-2$  $E_{\text{refl}} = 4.72\text{e}-2$  $E_{\text{refl}} = 5.34\text{e}-3$
Input    Face    Rim-spoke    Coarse-to-fine

Fig. 4. Coarse-to-fine deformation using different types of ARAP elements. The face-only ARAP elements sometimes result in a bumpy surface when the initial shape is not close to optimal (left). The rim-spoke elements are useful for finding the intermediate shape (middle) from which the face elements refine the solution more robustly (right).

### 4.4 Shape Updates

After a predefined number of descent steps, the vertex positions $\mathcal{V}'$ are updated from the optimized $\mathcal{T}$ by minimizing the ARAP-based energy $E_{\text{style}}$ using a standard local-global scheme [liu and Jacobson 2019; Liu and Jacobson 2021]. As suggested in the prior works, we pre-factorize the sparse linear system once for the reference mesh, and use it for the entire optimization. Please refer to [Liu and Jacobson 2021] for additional details on the shape update. Following the vertex update, $\mathcal{T}$ is re-initialized with the new face normals before entering the next gradient-descent cycle.

*Elements for ARAP energy.* Several types of elements can be used to define the ARAP-based energy in (3) with respect to the different trade-offs between robustness and deformation flexibility [Liu and Jacobson 2021]. For example, the rim-spoke element is known for its strong capability to preserve the original mesh; however, it prohibits the surface from forming sharp creases that are crucial for optimizing the stealth energy. In contrast, as suggested in [Liu and Jacobson 2021], the face-only element allows adjacent faces to rotate freely, whereas the mesh becomes more prone to becoming trapped in local minima. In our experiments, we employ a coarse-to-fine approach using the rim-spoke elements until the shape no longer exhibits significant deformation, and subsequently use the face-only elements to further reduce the reflectivity energy (Fig. 4).

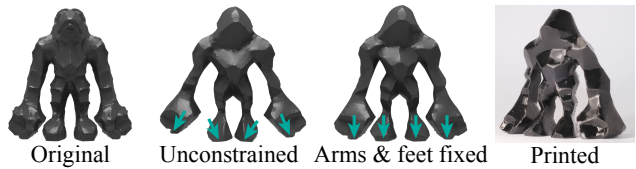

Original    Unconstrained    Arms & feet fixed    Printed

Fig. 5. Our ARAP-based deformation naturally incorporates positional constraints, such as fixing the feet of models such that they stand as in simulations after being printed.

*Positional constraints.* Constraints on the vertex positions can be easily incorporated into the gradient-based optimization by adding linear equality constraints to the ARAP step. This gives users a handle to control the deformation used to optimize the reflectivity, such as by fixing several vertex positions (Fig. 5).

### 4.5 Adaptive Edge Subdivision

Our workflow effectively balances reflectivity optimization and geometry preservation. However, one limitation of the ARAP-based energy is that it is bounded by mesh connectivity, which sometimes prevents the surface from forming a clean ridge. The formation of such a ridge is crucial for certain reflectivity objectives, such as the stealth energy. Although clean ridges could be obtained by restarting the optimization of $E_{\text{style}}$ [Zhao et al. 2017], we observed that maintaining the initial reference shape is crucial for the geometry preservation during reflectivity optimization. Thus, we address this issue by allowing the user to optionally run an adaptive mesh subdivision around problematic areas. For the subdivision operation, we focus on edge splits that introduce an additional bend at the midpoint of the selected edge. As the edge split does not change the geometry represented by the mesh (for example, edge flips can), the ARAP-based energy $E_{\text{style}}$ can be adjusted during optimization by performing the same operation on the reference mesh.

In remeshing, we consider two factors: the reflectivity-based criterion $C_{\text{refl}}$ and the geometric criterion $C_{\text{geom}}$. Our remeshing criterion is the product of these two criteria $C_{\text{refl}} C_{\text{geom}}$, which is activated when both criteria are sufficiently high. As the newly introduced faces incur additional computation for the reflectivity gradient estimation, which is costly compared to shape updates, it is desirable to limit the adaptivity to a few truly problematic regions. Accordingly, we select the edges of the top 5% in terms of the scalar remeshing criterion for splitting.

*Reflectivity-based criterion.* Because the local reflectivity energy indicates the potential to improve the total energy, it is reasonable to focus our search on faces with a high contribution $E_{\text{refl}}^k := A_k/|\mathcal{D}| \int_{\mathcal{D}} Q_k \, d\omega_l$. Thus, we define the reflectivity-based criterion using the one-ring of an edge $e$ (see the inset) as the sum $C_{\text{refl}}(e) := E_{\text{refl}}^{k_1} + E_{\text{refl}}^{k_2}$.

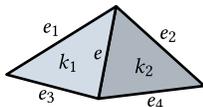

*Geometric criterion.* We assume the creases arising on the surface from our optimization to be smooth. Therefore, if two or more of the four neighboring edges of an edge $e$ have adjacent faces that significantly deviate from co-planar, it suggests that there is an undesirable corner in the ridge line, which we aim to fix using edge splitting. For example, if the edges $e_1$ and $e_2$ are bent by the optimization, a smoother ridge can be created by splitting $e$ and encouraging the ridge to move in the middle (Fig. 6). Thus, the geometric criterion is computed as $C_{\text{geom}}(e) := \sum_{i=1}^{4} |e_i| \arccos(\mathbf{n}_1^i \cdot \mathbf{n}_2^i)$. Here, $\mathbf{n}_1^i$ and $\mathbf{n}_2^i$ are the normals of the neighboring faces of $e_i$.

### 4.6 Implementation

We implemented our shape deformation method in C++ using libigl [Jacobson et al. 2018] for mesh processing algorithms. To evaluate the reflectivity energy and its derivatives, we implemented a path tracer and the radiative backpropagation [Nimier-David et al. 2020] using the Phong BRDF model [Phong 1975]. In our experiments, we sampled $N_{\text{path}} = 8$ paths up to the first indirect bounce to estimate radiance $L$, and independently generate $N_{\text{dir}} = 16$ directions for each face to evaluate the discretized equation (8). As the derivative can be computed in parallel on each face, we employed CPU multi-threading to accelerate our implementation.

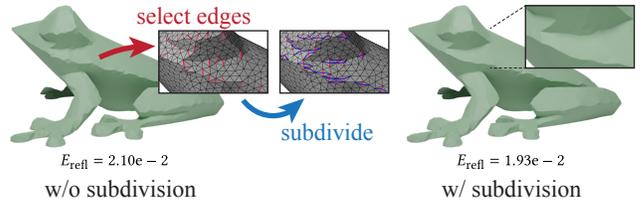

Fig. 6. The ARAP-based regularization results in several triangles not fully optimized (left). Our adaptive meshing finds problematic edges (shown in red) and then subdivides them (shown in blue) to further reduce $E_{\text{refl}}$ (right).

We set the number $n_{\text{gradient}}$ of gradient-based target normal updates before each vertex update to 8. We observed that the vertex update loop with each ARAP element type converged in approximately 30 iterations. Thus, we ran the vertex update for a fixed number of iterations to obtain the results. Specifically, we ran a coarse update with the rim-spoke element for 30 steps, and the face elements were used for the same number of steps. We then ran another 30 steps following the adaptive edge subdivision.

*Additional regularization.* It is often useful to employ a parameter that controls the strength of shape preservation. We could use the parameter $\lambda$ in the stylization energy (3), as used by Liu et al. [2019], to achieve varying degrees of alignment to target normals. However, our objective is to find a shape whose surface normals match the optimized targets in $\mathcal{T}$ as much as possible; therefore, we set $\lambda$ to a large constant ($\lambda = 1000$ in our experiments). Instead, we add a soft constraint to the target normals as $E_{\text{refl}} + \beta/2 \sum_k A_k \|\mathbf{t}_k - \mathbf{n}_k\|^2$ in the gradient-based reflectivity optimization step. Here, $\beta$ is a positive scalar, and $\mathbf{n}_k$ is the face normal of the original mesh. This additional regularization also alleviates the dependency of the results on other hyperparameters, such as $n_{\text{gradient}}$ and the step size $\eta$.

*Parameter values.* To obtain all the 3D-printed models, we used $\beta = 0.1$ and set the strength $\alpha$ of the total-variation denoising [Zhang et al. 2015b] to 250. Note that the denoising strength also depends on the scaling and the resolution of the input mesh. In our experiments, the input models were scaled such that the diagonal length of the bounding box is around 3. We used the step size $\eta = 200$. For creating Figs. 8, 7, 4, and 14, we performed minimal denoising ($\alpha = 1000$) to easily see the other parameter's effects.

## 5 RESULTS

We evaluate our framework by applying it to different reflectivity optimization problems. In Fig. 10, we optimize the stealth properties of the input 3D models and evaluate the results through simulations and real-world experiments. The fabricated shapes were 3D printed using a resin printer and painted with a glossy black spray. Fig. 11 displays the equipment used in the real-world measurement. We also demonstrate applications outside the stealth optimization (Fig. 12) and run a small ablation study illustrating our method's capability to find target normals to optimize reflection (Fig. 13).

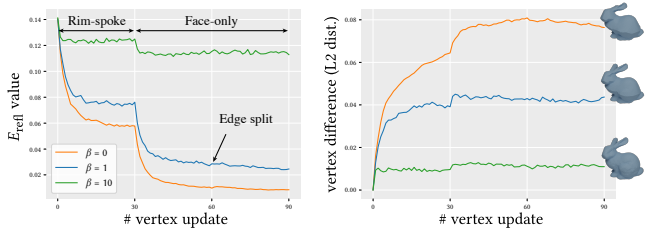

Fig. 7. The stages of convergence of our method using two types of ARAP elements, and then applying subdivision. We show three graphs with different strengths of the regularization parameter $\beta$. The left plot represents the reflectivity energy value, whereas the right plot shows the average vertex position difference to illustrate the shape preservation.

*Convergence analysis.* Fig. 7 shows the convergence behavior of the stealth optimization. The results suggest that our normal-driven optimization consistently reduces the energy with successive iterations, and specific techniques, such as face-only ARAP elements and adaptive edge splits, result in a smaller final energy.

*Comparison of regularization methods.* We qualitatively and quantitatively compare different gradient-based mesh deformation methods (Fig. 8). Directly modifying the vertex positions via the gradient descent results in a noisy surface with degraded mesh quality. Laplacian-based preconditioning [Nicolet et al. 2021] does not completely remove noise from the gradient, which leads to significant triangle distortions. In contrast, bi-Laplacian preconditioning alleviates the noise, but introduces a large surface stretch. Our method successfully preserves the original geometry and mesh quality during the optimization process.

*Timing analysis.* Table 1 lists the run-time performance results obtained by the proposed method. Each shape update iteration requires a few seconds for the models with 10K faces, allowing the user to interactively preview the intermediate results, adjust the parameters (e.g., the step size for gradient descent), or manually terminate the optimization when the desired trade-off between reflectivity and shape preservation is satisfied. The energy values over time is shown in Fig. 14. The performance bottleneck of the proposed algorithm is the gradient estimation, whose runtime scales almost linearly with the input mesh resolution. Although our implementation of the gradient estimation uses multi-threading on the CPU, we plan to accelerate this step further by employing GPU parallelism. We also plan to reduce the number of face elements using local face clustering [Cohen-Steiner et al. 2004].

Table 1. Performance breakdown of a single iteration in our optimization. We ran our algorithm on the BUNNY model with varying mesh resolutions. For the gradient estimation, we both report the time for a single gradient-based normal update and the total of 8 updates before each vertex step.

| # Faces | Grad. (single) | Grad. (total) | Denoise | Vertex | Total |
|---|---|---|---|---|---|
| 1K | 0.017s | 0.14s | 0.010s | 9.0e−3s | 0.18s |
| 10K | 0.23s | 1.94s | 0.095s | 0.17s | 2.5s |
| 50K | 1.3s | 10s | 1.0s | 0.65s | 13s |
| 100K | 2.6s | 21s | 1.4s | 2.7s | 28s |

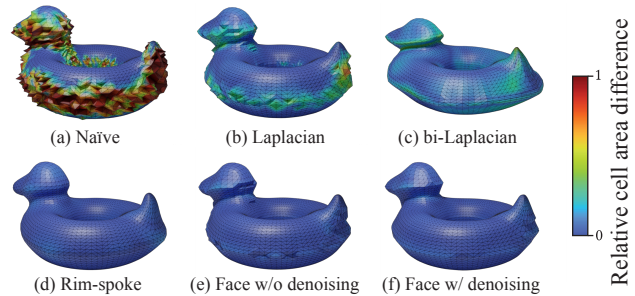

Fig. 8. Comparison of the different optimization strategies using reflectivity gradients. We visualize the relative cell area difference between the original and deformed shapes following the first 15 shape updates. (a) Directly updating vertex positions using noisy gradients. (b) Applying Laplacian preconditioning to the gradient. (c) Applying bi-Laplacian preconditioning. (d) Our deformation with the rim-spoke ARAP elements. (e) Our method with the face elements. (f) Our method with the face elements and denoising.

## 6 DISCUSSION AND CONCLUSION

We presented an algorithm that optimizes the geometry of a given 3D object to produce a shape with desirable reflection properties. We demonstrated the optimization of a 3D shape's stealth property with minimal deviation, and some other reflection or deflection requirements. However, our algorithm has several limitations.

First, our algorithm relies on a ray-based reflection model, wherein the wavelength is significantly smaller than the object's size. Although this is sufficient to reproduce the style of stealth craft, some of the actual designs of stealth aircraft solve a wave equation that considers surface details on the scale of the radar wavelength (a few centimeters) [Knott et al. 2004]. In addition, our algorithm uses gradient descent, which can be stacked in local minima.

In this study, we used ARAP-based regularization to encourage near-isometric deformation during reflectivity optimization. Although our regularization successfully preserves the original geometry in most cases, the isometric deformation can sometimes be too restrictive, leading to unpredictable shape optimization results (Fig. 9). In the future, we plan to consider other regularization approaches such as conformal deformation [Crane et al. 2013].

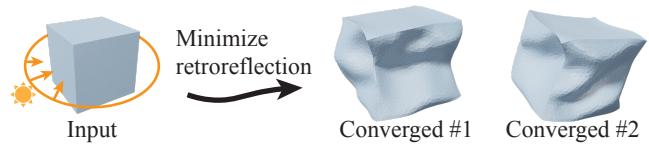

Fig. 9. Our reflectivity optimization does not always produce predictable results when the input shape is significantly far from an optimal shape.

Because of their simplicity and effectiveness, we believe that optimization-based modeling approaches such as ours can be incorporated into the design processes of objects and buildings. This can expand the designer's knowledge about the interaction of the design with light and the environment, and potentially even inspire new design forms that are not only aesthetic, but also follow certain functionalities as in Luis Sullivan's phrase.

## ACKNOWLEDGMENTS
The authors would like to thank Yuki Koyama and Takeo Igarashi for early discussions, and Yuta Yaguchi for support in 3D printing. This research is partially supported by the Israel Science Foundation grant number 1390/19.

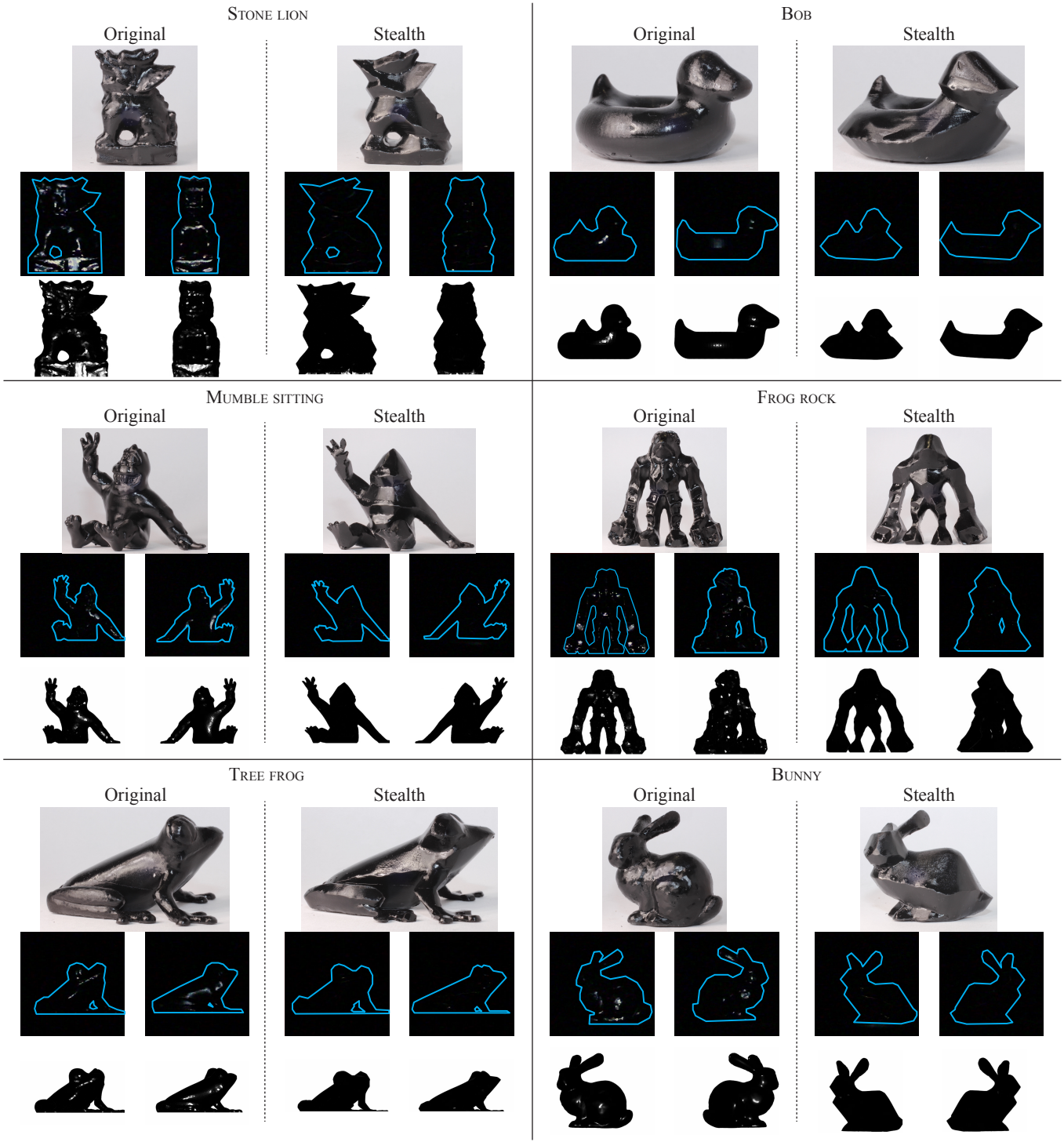

Fig. 10. Stealth optimization examples. Our method successfully minimizes the retroreflectivity of various 3D models, while preserving the overall geometry of the input. For each model, we compare the retroreflectivity of the original shape (left) with that of the shape optimized using our method (right) as seen from 2 different views. For each model, the middle row lists the reflection images obtained from real-world measurements, whereas the bottom row lists the simulation results.

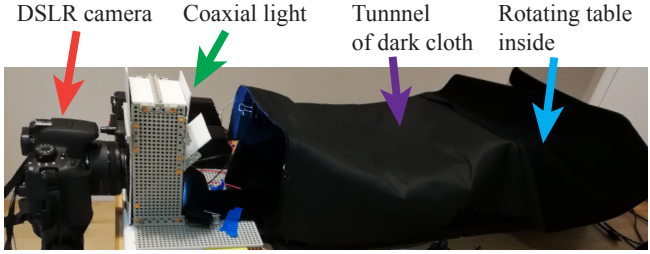

Fig. 11. Experiment setup. Parallel incoming and outgoing light directions assumed by the stealth energy were achieved by placing a half-mirror tilted 45° in front of the camera. The target object was placed on a rotating table.

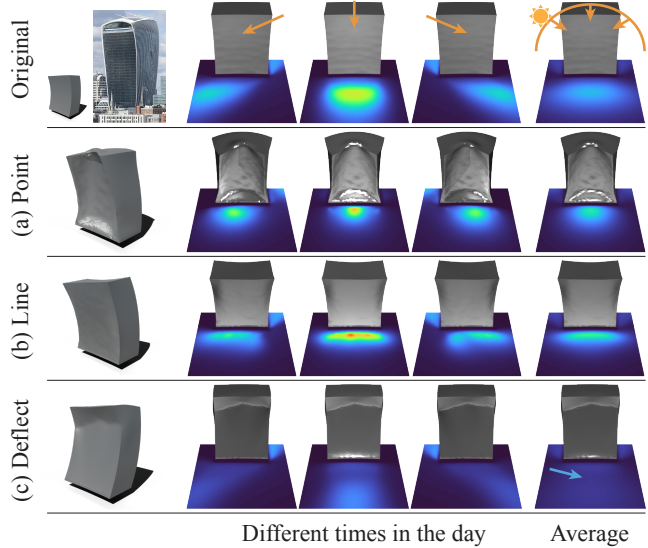

Fig. 12. Examples of reflectivity optimization enabled with our framework. We visualize the amount of sunlight reflection coming from the direction indicated by orange arrows. Our algorithm can find a building shape that collects sunlight (a) to a point or (b) on a line to optimize daylighting. Our method can also (c) deflect sunlight from a point (see blue arrow) to avoid undesirable sunlight concentration [Zhu et al. 2019]. Examples (a) and (b) increased the energy respectively to 350% and 280% of the input shape's value, and example (c) decreased the energy to the original's 9%. The image is from wikipedia.com by Colin under CC BY-SA 4.0.

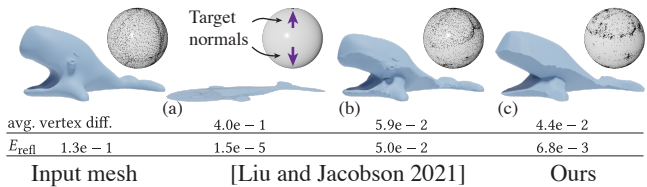

| | (a) | (b) | (c) |
|---|---|---|---|
| avg. vertex diff. | 4.0e − 1 | 5.9e − 2 | 4.4e − 2 |
| $E_{\text{refl}}$  1.3e − 1 | 1.5e − 5 | 5.0e − 2 | 6.8e − 3 |
| Input mesh | [Liu and Jacobson 2021] | | Ours |

Fig. 13. Comparison with the approach by Liu et al. [2021]. Next to each model, we visualize its face normals as the points on a sphere. Fixed target normals of ±z directions either (a) completely crush the mesh, or (b) leave a significant portion of unoptimized normals, when used with a weaker normal alignment ($\lambda = 2.0e − 3$). In contrast, (c) our method effectively finds target normals that optimize the reflectivity, resulting in significantly lower $E_{\text{refl}}$ with smaller shape distortion.

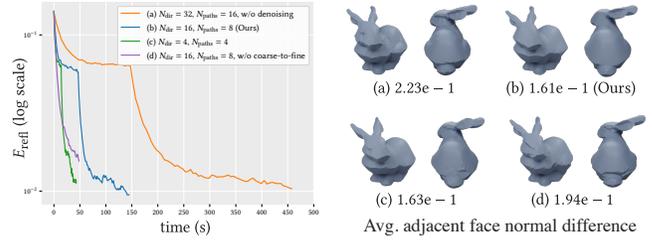

Fig. 14. Reflectivity energy values over time (left) and the average difference in adjacent face normals to illustrate the significance of noise in the final shape (right). The model has 6K faces. Naïvely running our algorithm without denoising (a) typically introduces noise even with relatively large sampling counts. Our setting (b) uses lower sampling counts and denoising, significantly improving both runtime and convergence. Our framework also provides several *preview modes*, such as (c) using even smaller sampling counts, or (d) only running the fine step with the face elements, which may introduce additional noise, but accelerate the feedback cycles to rapidly explore desirable parameter values. Note that, although the difference between (b) and (c) is subtle in the numbers, bumps introduced by noise are often difficult to remove afterward in our optimization.